\documentclass[journal,twoside]{IEEEtran}
\usepackage{amsmath}
\usepackage{amssymb}
 \usepackage{amsthm}
 \usepackage{hyperref}
 \usepackage{subfigure}
 \usepackage{tikz}
 \usetikzlibrary{arrows,positioning,shadows}
 \usepackage{standalone}
 \usepackage{algorithmic}
 \usepackage{algorithm}
 \usepackage{tabularx,ragged2e,booktabs,caption}
 \usepackage{flexisym}
 \usepackage{pseudocode}
 \usepackage{multirow}

\begin{document}

\newcounter{RZNumberOfComments}
\stepcounter{RZNumberOfComments}

\newcommand{\reza}[1]{\textcolor{red}{\small \bf [REZA\#\arabic{RZNumberOfComments}\stepcounter{RZNumberOfComments}: #1]}}
\newcommand{\amir}[1]{\textcolor{blue}{\small \bf [AMIR\#\arabic{RZNumberOfComments}\stepcounter{RZNumberOfComments}: #1]}}

\title{NetSpam: a Network-based Spam Detection Framework for Reviews in Online Social Media}

\author{Saeedreza Shehnepoor, Mostafa Salehi*, Reza Farahbakhsh, Noel Crespi
\thanks{S.R. Shehnepoor is with the University of Tehran, Tehran, Iran.
M. Salehi (*corresponding author) is with the University of Tehran, Tehran, Iran.
R. Farahbakhsh is with the Institut Mines-Telecom, Telecom SudParis, Paris, France.
N. Crespi is with the Institut Mines-Telecom, Telecom SudParis, Paris, France.
emails: \{shehnepoor@ut.ac.ir, mostafa\_salehi@ut.ac.ir, reza.farahbakhsh@it-sudparis.eu, noel.crespi@institut-telecom.fr.\}}
}

%%%%%%%%%%%%%%%%%%%%%%%%%%%%%%%%%%%%%%%%%%%%%%%%%
%\input{0_Abstract}
\IEEEtitleabstractindextext{

\begin{abstract}
Nowadays, a big part of people rely on available content in social media in their decisions (e.g. reviews and feedback on a topic or product). The possibility that anybody can leave a review provide a golden opportunity for spammers to write spam reviews about products and services for different interests. Identifying these spammers and the spam content is a hot topic of research and although a considerable number of studies have been done recently toward this end, but so far the methodologies put forth still barely detect spam reviews, and none of them show the importance of each extracted feature type.
In this study, we propose a novel framework, named \textsl{NetSpam}, which utilizes spam features for modeling review datasets as heterogeneous information networks to map spam detection procedure into a classification problem in such networks. Using the importance of spam features help us to obtain better results in terms of different metrics experimented on real-world review datasets from Yelp and Amazon websites.
The results show that \textit{NetSpam} outperforms the existing methods and among four categories of features; including review-behavioral, user-behavioral, review-linguistic, user-linguistic, the first type of features performs better than the other categories.
\end{abstract}

}

\maketitle

\IEEEdisplaynontitleabstractindextext

%%%%%%%%%%%%%%%%%%%%%%%%%%%%%%%%%%%%%%%%%
%\input{1_Introduction}
\section{Introduction}
Online Social Media portals play an influential role in information propagation which is considered as an important source for producers in their advertising campaigns as well as for customers in selecting products and services.
In the past years, people rely a lot on the written reviews in their decision-making processes, and positive/negative reviews encouraging/discouraging them in their selection of products and services. 
In addition, written reviews also help service providers to enhance the quality of their products and services.
These reviews thus have become an important factor in success of a business while positive reviews can bring benefits for a company, negative reviews can potentially impact credibility and cause economic losses. The fact that anyone with any identity can leave comments as review, provides a tempting opportunity for spammers to write fake reviews designed to mislead users' opinion. These misleading reviews are then multiplied by the sharing function of social media and propagation over the web. The reviews written to change users' perception of how good a product or a service are considered as spam \cite{Akoglu2013}, and are often written in exchange for money. 
As shown in \cite{yelpRate}, 20\% of the reviews in the Yelp website are actually spam reviews.
 
On the other hand, a considerable amount of literature has been published on the techniques used to identify spam and spammers as well as different type of analysis on this topic \cite{Heydari2015, Crawford2015}.
These techniques can be classified into different categories; some using linguistic patterns in text \cite{Ott2012}, \cite{Ott2011}, \cite{Xu2014}, which are mostly based on bigram, and unigram, others are based on behavioral patterns that rely on features extracted from patterns in users' behavior which are mostly metadata-based \cite{Jindal2008,Li2011,Fei2013,Minnich2015}, \cite{Viswanath2014}, and even some techniques using graphs 
and graph-based algorithms and classifiers \cite{Li2014,Akoglu2013,Shebuti2015}.

Despite this great deal of efforts, many aspects have been missed or remained unsolved. One of them is a classifier that can calculate feature weights that show each feature's level of importance in determining spam reviews. 
The general concept of our proposed framework is to model a given review dataset as a Heterogeneous Information Network (HIN) \cite{Sun2012} and to map the problem of spam detection into a HIN classification problem. 
In particular, we model review dataset as a HIN in which reviews are connected through different node types (such as features and users). A weighting algorithm is then employed to calculate each feature's importance (or weight). These weights are utilized to calculate the final labels for reviews using both unsupervised and supervised approaches. 

To evaluate the proposed solution, we used two sample review datasets from Yelp and Amazon websites. 
Based on our observations, defining two views for features (review-user and behavioral-linguistic), the classified features as review-behavioral have more weights and yield better performance on spotting spam reviews in both semi-supervised and unsupervised approaches. In addition, we demonstrate that using different supervisions such as 1\%, 2.5\% and 5\% or using an unsupervised approach, make no noticeable variation on the performance of our approach. We observed that feature weights can be added or removed for labeling and hence time complexity can be scaled for a specific level of accuracy.
As the result of this weighting step, we can use fewer features with more weights to obtain better accuracy with less time complexity.
In addition, categorizing features in four major categories (review-behavioral, user-behavioral, review-linguistic, user-linguistic), helps us to understand how much each category of features is contributed to spam detection. 

In summary, our main contributions are as follows:

(i)
We propose \textsl{NetSpam} framework that is a novel network-based approach which models review networks as heterogeneous information networks.
The classification step uses different metapath types which are innovative in the spam detection domain.

(ii)
A new weighting method for spam features is proposed to determine the relative importance of each feature and shows how effective each of features are in identifying spams from normal reviews. 
Previous works \cite{Shebuti2015, Mukherjee2013-2} also aimed to address the importance of features mainly in term of obtained accuracy, but not as a build-in function in their framework (i.e., their approach is dependent to ground truth for determining each feature importance). As we explain in our unsupervised approach, \textsl{NetSpam} is able to find features importance even without ground truth, and only by relying on metapath definition and based on values calculated for each review.

(iii)
\textsl{NetSpam} improves the accuracy compared to the state-of-the art in terms of time complexity, which highly depends to the number of features used to identify a spam review; hence, using features with more weights will resulted in detecting fake reviews easier with less time complexity.

%%%%%%%%%%%%%%%%%%%%%%%%%%%%%%%%%%%%%%%%%%%
%\input{3_Preliminaries}
\section{Preliminaries}
\label{sec:Preliminaries}

As mentioned earlier, we model the problem as a heterogeneous network where nodes are either real components in a dataset (such as reviews, users and products) or spam features.
To better understand the proposed framework we first present an overview of some of the concepts and definitions in heterogeneous information networks \cite{Sun2009}, \cite{Sun2011}, \cite{Chen2014}.

\subsection{Definitions}
\label{definitions}
\textbf{Definition 1 (Heterogeneous Information Network).} Suppose we have $r (>1)$ types of nodes and $s (>1)$ types of relation links between the nodes, then a heterogeneous information network is defined as a graph $G = (V,E)$ where each node $v\in V$ and each link $e\in E$ belongs to one particular node type and link type respectively. If two links belong to the same type, the types of starting node and ending node of those links are the same. 

\textbf{Definition 2 (Network Schema).} Given a heterogeneous information network $G=(V,E)$, a network schema $T = (A,R)$ is a metapath with the object type mapping $\tau: V \rightarrow A$ and link mapping $\phi: E \rightarrow R$, which is a graph defined over object type $A$, with links as relations from $R$. The schema describes the metastructure of a given network (i.e., how many node types there are and where the possible links exist).

\textbf{Definition 3 (Metapath).} As mentioned above, there are no edges between two nodes of the same type, but there are paths. Given a heterogeneous information network $G =(V,E)$, a metapath $P$ is defined by a sequence of relations in the network schema $T = (A,R)$, denoted in the form $A_1(R_1)A_2(R_2)...(R_{(l-1)})A_l$, which defines a composite relation $P = R_1  o R_2  o...o R_{(l-1)}$ between two nodes, where $o$ is the composition operator on relations. For convenience, a metapath can be represented by a sequence of node types when there is no ambiguity, i.e., $P=A_1 A_2...A_l$. The metapath extends the concept of link types to path types and describes the different relations among node types through indirect links, i.e. paths, and also implies diverse semantics.

\textbf{Definition 4 (Classification problem in heterogeneous information networks).}  Given a heterogeneous information network $G = (V,E)$, suppose $V\textprime$ is a subset of V that contains nodes of the target type (i.e., the type of nodes to be classified). $k$ denotes the number of the class, and for each class, say $C_1...C_k $, we have some pre-labeled nodes in $V\textprime$ associated with a single user. The classification task is to predict the labels for all the unlabeled nodes in $V\textprime$.

\subsection{Feature Types}
\label{features-type}
In this paper, we use an extended definition of the metapath concept as follows.
A metapath is defined as a path between two nodes, which indicates the connection of two nodes through their shared features.
When we talk about metadata, we refer to its general definition, which is data about data. In our case, the data is the written review, and by metadata we mean data about the reviews, including user who wrote the review, the business that the review is written for, rating value  of the review, date of written review and finally its label as spam or genuine review.

In particular, in this work features for users and reviews fall into the categories as follows (shown in Table \ref{Feature-list}):

\textbf{Review-Behavioral (RB) based features.} This feature type is based on metadata and not the review text itself. 
The RB category contains two features; Early time frame (ETF) and Threshold rating deviation of review (DEV) \cite{Mukherjee2013-1}.

\textbf{Review-Linguistic (RL) based features.} \textcolor[rgb]{0,0,0}{Features in this category are based on the review itself and extracted directly from text of the review. In this work we use two main features in RL category; the Ratio of 1st Personal Pronouns (PP1) and the Ratio of exclamation sentences containing `!' (RES) \cite{Li2011}.}

\textbf{User-Behavioral (UB) based features.} These features are specific to each individual user and they are calculated per user, so we can use these features to generalize all of the reviews written by that specific user. This category has two main features; the Burstiness of reviews written by a single user \cite{Fei2013}, and the average of a users' negative ratio given to different businesses \cite{Mukherjee2013-2}.

\textbf{User-Linguistic (UL) based features.} These features are extracted from the users' language and shows how users are describing their feeling or opinion about what they've experienced as a customer of a certain business. 
We use this type of features to understand how a spammer communicates in terms of wording. There are two features engaged for our framework in this category; Average Content Similarity (ACS) and Maximum Content Similarity (MCS). These two features show how much two reviews written by two different users are similar to each other, as spammers tend to write very similar reviews by using template pre-written text \cite{Akoglu2013}.

\begin{table*}[t]
\centering
\caption {Features for users and reviews in four defined categories (the calculated values are based on Table 2 in \cite{Shebuti2015})} 
\vspace{-0.2cm}
\label{Feature-list}
\begin{tabular}{|>{\centering\arraybackslash}m{1.2cm}|>{\arraybackslash}m{6.4cm}|>{\arraybackslash}m{9cm}|}\hline
{Spam Feature}&{User-based}&{Review-based} \\ \hline
{Behavioral-based Features}&{\textit{Burstiness} \cite{Mukherjee2013-2}: Spammers, usually write their spam reviews in short period of time for two reasons: first, because they want to impact readers and other users, and second because they are temporal users, they have to write as much as reviews they can in short time.
\begin{align}
x_{BST}(i)=
\begin{cases}
0 \quad  (L_{i}-F_{i}) \notin (0,\tau)\\
1-\frac{L_{i}-F_{i}}{\tau} \quad (L_{i}-F_{i}) \in (0,\tau)\\
\end{cases}
\end{align}
where $L_i - F_i$ describes days between last and first review for $\tau = 28$.
 Users with calculated value greater than 0.5 take value 1 and others take 0.
\newline

\textit{Negative Ratio} \cite{Mukherjee2013-2}: Spammers tend to write reviews which defame businesses which are competitor with the ones they have contract with, this can be done with destructive reviews, or with rating those businesses with low scores. Hence, ratio of their scores tends to be low. Users with average rate equal to 2 or 1 take 1 and others take 0. }&{\textit{Early Time Frame} \cite{Mukherjee2013-1}: Spammers try to write their reviews asap, in order to keep their review in the top reviews which other users visit them sooner.
\begin{align}
x_{ETF}(i)=
\begin{cases}
0 \quad  (T_{i}-F_{i}) \notin (0,\delta)\\
1-\frac{T_{i}-F_{i}}{\delta}\quad  (T_{i}-F_{i}) \in (0,\delta)\\
\end{cases}
\end{align}
where $L_i - F_i$ denotes days specified written review and first written review for a specific business. We have also $\delta = 7$.
Users with calculated value greater than 0.5 takes value 1 and others take 0.
\newline

\textit{Rate Deviation using threshold} \cite{Mukherjee2013-1}: Spammers, also tend to promote businesses they have contract with, so they rate these businesses with high scores. In result, there is high diversity in their given scores to different businesses which is the reason they have high variance and deviation.
\begin{align}
x_{DEV}(i)=
\begin{cases}
0 \quad  otherwise\\
1-\frac{r_{ij}-avg_{e\in E_{*j}}r(e)}{4}>\beta_1\\
\end{cases}
\end{align}
where $\beta_{1}$ is some threshold determined by recursive minimal entropy partitioning.
  Reviews are close to each other based on their calculated value, take same values (in $[0,1)$).} \\ 
\hline
{Linguistic-based Features}&{\textit{Average Content Similarity} \cite{Fei2013}, \textit{Maximum Content Similarity} \cite{Mukherjee2013-1}: \textcolor[rgb]{0,0,0}{Spammers, often write their reviews with same template and they prefer not to waste their time to write an original review. In result, they have similar reviews.} Users have close calculated values take same values (in $[0,1)$).}&{\textit{Number of first Person Pronouns}, \textit{Ratio of Exclamation Sentences containing `!'} \cite{Li2011}: First, studies show that spammers use second personal pronouns much more than first personal pronouns. In addition, spammers put '!' in their sentences as much as they can to increase impression on users and highlight their reviews among other ones. Reviews are close to each other based on their calculated value, take same values (in $[0,1)$).} \\ 
\hline
\end{tabular}
\end{table*}

%%%%%%%%%%%%%%%%%%%%%%%%%%%%%%%%%%%%%%%%%%%%
%\input{4_NetSpam}
\section{NetSpam; The Proposed Solution}
\label{sec:ProposedMeasurmentProcess}

In this section, we provides details of the proposed solution which is shown in Algorithm \ref{NetSpam-algorithm}.

\subsection{Prior Knowledge}
\label{sec:review-network-section}

The first step is computing prior knowledge, i.e. the initial probability of review $u$ being spam which denoted as $y_u$. The proposed framework works in two versions; semi-supervised learning and unsupervised learning. In the semi-supervised method, $y_u=1$ if review $u$ is labeled as spam in the pre-labeled reviews, otherwise $y_u=0$. If the label of this review is unknown due the amount of supervision, we consider $y_u=0$ (i.e., we assume $u$ as a non-spam review). In the unsupervised method, our prior knowledge is realized by using
$y_u = (1/L)\sum_{l=1}^{L}f(x_{lu})$ where $f(x_{lu})$ is the probability of review $u$ being spam according to feature $l$ and $L$ is the number of all the used features (for details, refer to \cite{Shebuti2015}).

\subsection{Network Schema Definition}
\begin{figure}
\begin{center}
\includegraphics[width=7cm]{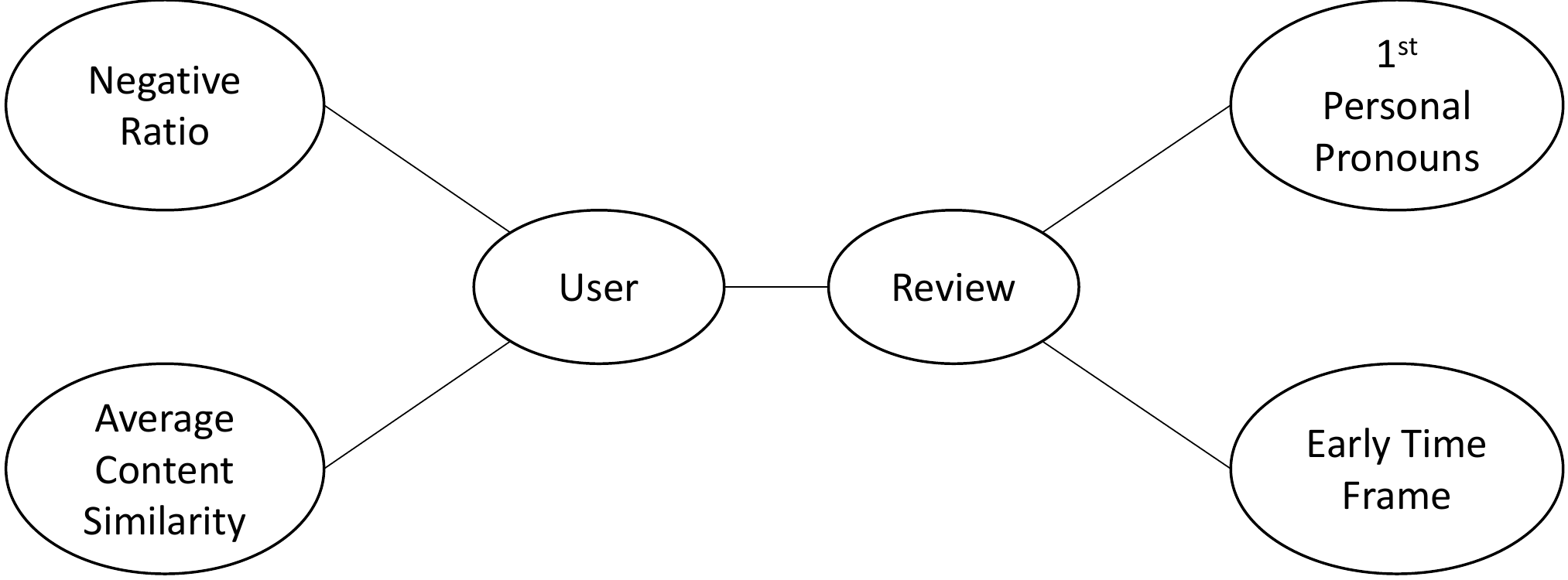}
\vspace{-0.2cm}
\caption {An example for a network schema generated based on a given spam features list; NR, ACS, PP1 and ETF.}
\label{fig:schemes}
\end{center}
\vspace{-0.35cm}
\end{figure}

The next step is defining network schema based on a given list of spam features which determines the features engaged in spam detection. This Schema are general definitions of metapaths and show in general how different network components are connected. For example, if the list of features includes NR, ACS, PP1 and ETF, the output schema is as presented in Fig. \ref{fig:schemes}.

\subsection{Metapath Definition and Creation}
\label{metapath-creation}
As mentioned in Section \ref{definitions}, a metapath is defined by a sequence of relations in the network schema. 
Table \ref{MetaPaths} shows all the metapaths used in the proposed framework. As shown, the length of user-based metapaths is 4 and the length of review-based metapaths is 2.

For metapath creation, we define an extended version of the metapath concept considering different levels of spam certainty. In particular, two reviews are connected to each other if they share same value. Hassanzadeh \textit{et al.} \cite{Hassanzadeh2014} propose a fuzzy-based framework and indicate for spam detection, it is better to use fuzzy logic for determining a review's label as a spam or non-spam. Indeed, there are different levels of spam certainty. 
We use a step function to determine these levels. 
In particular, given a review $u$, the levels of spam certainty for metapath $p_l$ (i.e., feature $l$) is calculated as $m_u^{p_l} = \frac{\lfloor s \times  f(x_{lu})\rfloor}{s}$, where $s$ denotes the number of levels. 
After computing $m_u^{p_l}$ for all reviews and metapaths, two reviews $u$ and $v$ with the same metapath values (i.e., $m_u^{p_l}=m_v^{p_l}$) for metapath $p_l$ are connected to each other through that metapath and create one link of review network. The metapath value between them denoted as $m_{u,v}^{p_l}=m_u^{p_l}$.

Using $s$ with a higher value will increase the number of each feature's metapaths and hence fewer reviews would be connected to each other through these features. Conversely, using lower value for $s$ leads us to have bipolar values (which means reviews take value 0 or 1).
\textcolor[rgb]{0,0,0}{Since we need enough spam and non-spam reviews for each step, with fewer number of reviews connected to each other for every step, the spam probability of reviews take uniform distribution, but with lower value of $s$ we have enough reviews to calculate final spamicity for each review. Therefore, accuracy for lower levels of $s$ decreases because of the bipolar problem, and it decades for higher values of $s$, because they take uniform distribution. In the proposed framework, we considered $s = 20$,  i.e. $m_u^{p_l} \in \{0,0.05,0.10,...,0.85,0.90,0.95\}$.} 

\begin{minipage}{0.5\columnwidth}
\vspace{-0.2cm}
\begin{pseudocode}[ruled] {NetSpam}{\null}
  \label{NetSpam-algorithm}
	Input: review-dataset, spam-feature-list,\\
		 pre-labeled-reviews\\
	Output: features-importance (W),\\
		 spamicity-probability (Pr) \\
	\mbox{\textcolor[rgb]{0,0,0}{\% $u,v$: review, $y_u$: spamicity probability of review $u$}} \\
	\mbox{\% $f(x_{lu})$: initial probability of review $u$ being spam} \\	
	\mbox{\textcolor[rgb]{0,0,0}{\% $p_l$: metapath based on feature $l$, $L$: features number}}\\
	\mbox{\textcolor[rgb]{0,0,0}{\% $n$: number of reviews connected to a review}} \\
	\mbox{\textcolor[rgb]{0,0,0}{\% $m_u^{p_l}$: the level of spam certainty}} \\
	\mbox{\textcolor[rgb]{0,0,0}{\% $m_{u,v}^{p_l}$: the metapath value}} \\
%	\mbox{\textcolor[rgb]{0,0,0}{\% m: metapath value, mp: metapth}} \\
	\mbox{\%Prior Knowledge}\\
	\IF \mbox{semi-supervised mode} \\
				\BEGIN
				\IF{\mbox{$u \in pre-labeled-reviews$}} \\
				\BEGIN
					\mbox{$y_u = label(u)$}\\
				\END
				\ELSE  \\
				\BEGIN
					\mbox{$y_u = 0$} \\
				\END
				\END
			\ELSE  \mbox{\% unsupervised mode} \\
			\BEGIN
				\mbox{$y_u = \frac{1}{L}\sum_{l=1}^{L}f(x_{lu})$} \\
			\END \\
\mbox{ \%Network Schema Definition} \\
				\mbox{$schema$ = defining schema based on spam-feature-list}\\
	\mbox{ \% Metapath Definition and Creation} \\
	\FOR {p_l \in schema} \DO
	\BEGIN	
	\FOR {u,v \in review-dataset} \DO
		\BEGIN
			\mbox{$m_u^{p_l} = \frac{\lfloor s \times  f(x_{lu})\rfloor}{s}$}\\
			\mbox{$m_v^{p_l} = \frac{\lfloor s \times  f(x_{lv})\rfloor}{s}$}\\
		\IF {m_u^{p_l} = m_v^{p_l}} \\
			\BEGIN
				\mbox{$mp_{u,v}^{p_l} = m_u^{p_l}$} \\
			\END
			\ELSE \\
			\BEGIN
				\mbox{$mp_{u,v}^{p_l} = 0$} \\
			\END
	\END \\
	\END \\
	\mbox{\% Classification - Weight Calculation} \\
	\FOR {p_l \in schemes} \DO
		\BEGIN	
			\mbox{$W_{p_l} = \frac{\sum_{r=1}^{n}\sum_{s=1}^{n}mp_{r,s}^{p_l}\times y_{r}\times y_{s}}{\sum_{r=1}^{n}\sum_{s=1}^{n}mp_{r,s}^{p_l}} $} \\
		\END \\
	\mbox{\% Classification - Labeling} \\
	\FOR {u,v \in review-dataset} \DO
		\BEGIN
			\mbox{$Pr_{u,v} = 1 - \Pi_{p_l=1}^{L} {1-{mp_{u,v}^{p_l}\times W_{p_l}}}$}\\ [0.2in]
			\mbox{$Pr_{u}= avg(Pr_{u,1},Pr_{u,2},...,Pr_{u,n})$} \\	
		\END \\
	\RETURN {\mbox{W, Pr}}
\end{pseudocode}
\end{minipage}

\begin{table*}
\begin{center}
\caption {Metapaths used in the \textsl{NetSpam} framework.}
\vspace{-.2cm}
\label{MetaPaths}
\begin{tabular}{|>{\centering\arraybackslash}m{0.5cm}|>{\centering\arraybackslash}m{1.8cm}|>{\centering\arraybackslash}m{0.5cm}|>{\centering\arraybackslash}m{5cm}|>{\centering\arraybackslash}m{7.5cm}|}\hline
{\textbf{Row}}&{\textbf{Notation}}&{\textbf{Type}}&{\textbf{MetaPath}}&{\textbf{Semantic}} \\ \hline
%{1}&{R-U-R}&{Review-User-Review}&{Reviews written  by same User} \\ \hline
{1}&{R-DEV-R}&{RB}&{Review-Threshold Rate Deviation-Review}&{Reviews with same Rate Deviation from average Item rate (based on recursive minimal entropy partitioning)} \\ \hline
{2}&{R-U-NR-U-R}&{UB}&{Review-User-Negative Ratio-User-Review}&{Reviews written by different Users with same Negative Ratio} \\ \hline
{3}&{R-ETF-R}&{RB}&{Review-Early Time Frame-Review}&{Reviews with same released date related to Item} \\ \hline
{4}&{R-U-BST-U-R}&{UB}&{Review-User-Burstiness-User-Review}&{Reviews written  by different users in same Burst} \\ \hline
{5}&{R-RES-R}&{RL}&{Review-Ratio of Exclamation Sentences containing `!'-Review}&{Reviews with same number of Exclamation Sentences containing `!'} \\ \hline
{6}&{R-PP1-R}&{RL}&{Review-first Person Pronouns-Review}&{Reviews with same number of first Person Pronouns} \\ \hline
{7}&{R-U-ACS-U-R}&{UL}&{Review-User-Average Content Similarity-User-Review}&{Reviews written  by different Users with same Average Content Similarity using cosine similarity score} \\ \hline
{8}&{R-U-MCS-U-R}&{UL}&{Review-User-Maximum Content Similarity-User-Review}&{Reviews written  by different Users with same Maximum Content Similarity using cosine similarity score} \\ \hline
\end{tabular}	
\end{center}
\end{table*}

\subsection{Classification}
The classification part of \textit{NetSpam} includes two steps; (i) \textsl{weight calculation} which determines the importance of each spam feature in spotting spam reviews, (ii) \textsl{Labeling} which calculates the final probability of each review being spam. Next we describe them in detail.

\subsubsection{Weight Calculation}
This step computes the weight of each metapath. We assume that nodes' classification is done based on their relations to other nodes in the review network; linked nodes may have a high probability of taking the same labels.
The relations in a heterogeneous information network not only include the direct link but also the path that can be measured by using the metapath concept. Therefore, we need to utilize the metapaths defined in the previous step, which represent heterogeneous relations among nodes. 
Moreover, this step will be able to compute the weight of each relation path (i.e., the importance of the metapath), which will be used in the next step (Labeling) to estimate the label of each unlabeled review.

The weights of the  metapaths will answer an important question; which metapath (i.e., spam feature) is better at ranking spam reviews? 
Moreover, the weights help us to understand the formation mechanism of a spam review. In addition, since some of these spam features may incur considerable computational costs (for example, computing linguistic-based features through $NLP$ methods in a large review dataset), choosing the more valuable features in the spam detection procedure leads to better performance whenever the computation cost is an issue.

To compute the weight of metapath $p_i$, for $i=1,...,L$ where $L$ is the number of metapaths, we propose following equation:
\begin{equation}
W_{p_i} = \frac{\sum_{r=1}^{n}\sum_{s=1}^{n}mp_{r,s}^{p_i}\times y_{r}\times y_{s}}{\sum_{r=1}^{n}\sum_{s=1}^{n}mp_{r,s}^{p_i}}
\end{equation}
where $n$ denotes the number of reviews and $mp_{r,s}^{p_i}$ is a metapath value between reviews $r$ and $s$ if there is a path between them through metapath $p_i$, otherwise $mp_{r,s}^{p_i}=0$. Moreover, $y_{r} (y_{s})$ is 1 if review $r (s)$ is labeled as spam in the pre-labeled reviews, otherwise 0.

\subsubsection{Labeling}
\begin{figure*}
\includegraphics[width=\textwidth,height=8cm]{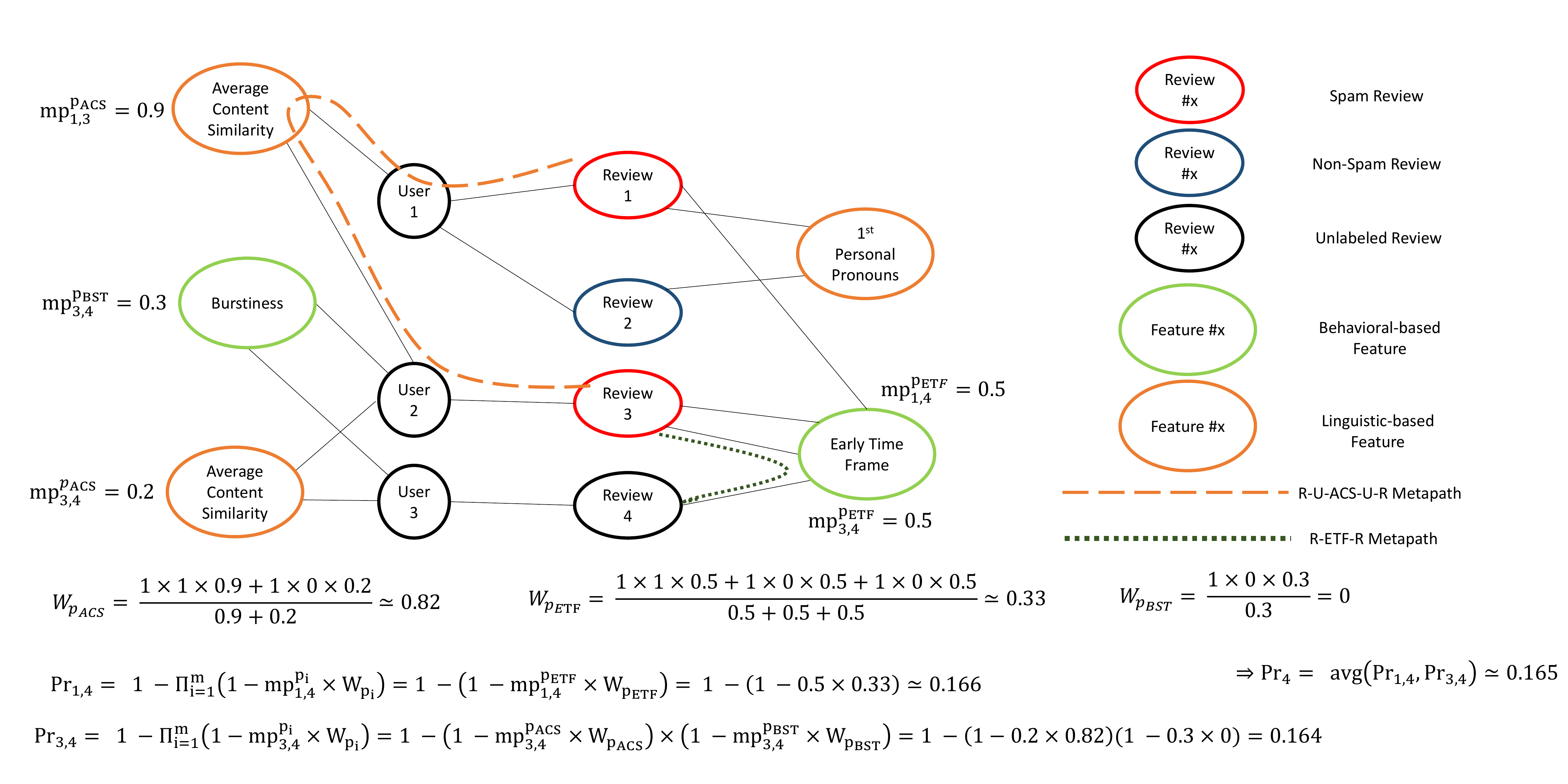}
\caption {An example of a review network and different steps of proposed framework.}
\label{fig:metapath}
\end{figure*}

Let $Pr_{u,v}$ be the probability of unlabeled review $u$ being spam by considering its relationship with spam review $v$.  To estimate $Pr_{u}$, the probability of unlabeled review $u$ being spam, we propose the following equations:

\begin{equation}
Pr_{u,v} = 1 - \Pi_{i=1}^{L} {1-mp_{u,v}^{p_i}\times W_{p_i}}
\end{equation}

\begin{equation}
Pr_{u} = avg(Pr_{u,1},Pr_{u,2},...,Pr_{u,n})
\end{equation}
where $n$ denotes number of reviews connected to review $u$.
Fig. \ref{fig:metapath} shows an example of a review network and different steps of proposed framework.

\textcolor[rgb]{0,0,0}{
It is worth to note that in creating the HIN, as much as the number of links between a review and other reviews increase, its probability to have a label similar to them increase too, because it assumes that a node relation to other nodes show their similarity. In particular, more links between a node and other non-spam reviews, more probability for a review to be non-spam and vice versa. In other words, if a review has lots of links with non-spam reviews, it means that it shares features with other reviews with low spamicity and hence its probability to be a non-spam review increases.
} 

%%%%%%%%%%%%%%%%%%%%%%%%%%%%%%%%%%%%%%%%%%%
%\input{5_Evaluation}
\section{Experimental Evaluation}
\label{sec:expEval}
\textcolor[rgb]{0,0,0}{This section presents the experimental evaluation part of this study including the datasets and the defined metrics as well as the obtained results.}

\subsection{Datasets}
\textcolor[rgb]{0,0,0}{
Table \ref{Dataset} includes a summary of the datasets and their characteristics.
We used a dataset from Yelp, introduced in \cite{Shebuti2015}, which includes almost 608,598 reviews written by customers of restaurants and hotels in NYC. 
The dataset includes the reviewers' impressions and comments about the quality, and other aspects related to a restaurants (or hotels). 
The dataset also contains labeled reviews as ground truth (so-called near ground-truth \cite{Shebuti2015}), which indicates whether a review is spam or not.
Yelp dataset was labeled using filtering algorithm engaged by the Yelp recommender, and although none of recommenders are perfect, but according to \cite{yelpRecomm} it produces trustable results. It explains hiring someone to write different fake reviews on different social media sites, it is the yelp algorithm that can spot spam reviews and rank one specific spammer at the top of spammers.
Other attributes in the dataset are rate of reviewers, the date of the written review, and date of actual visit, as well as the user's and the restaurant's id (name). 
}

We created three other datasets from this main dataset as follow:

\textsl{- Review-based} dataset, includes 10\% of the reviews from the \textsl{Main} dataset, randomly selected using uniform distribution.

\textsl{- Item-based} dataset, composes of 10\% of the randomly selected reviews of each item, also based on uniform distribution (as with Review-based dataset). %\reza{what means this?}.

\textsl{- User-based} dataset, includes randomly selected reviews using uniform distribution in which one review is selected from every 10 reviews of single user and if number of reviews was less than 10, uniform distribution has been changed in order to at least one review from every user get selected.

\textcolor[rgb]{0,0,0}{In addition to the presented dataset, we also used another real-world set of data from Amazon \cite{Jindal2008} to evaluate our work on unsupervised mode. There is no credible label in the Amazon dataset (as mentioned in \cite{Mukherjee2016}), but we used this dataset to show how much our idea is viable on other datasets beyond Yelp and results for this dataset is presented on Sec. \ref{sec:unsupervised}.}

\begin{table}[t]
\vspace{-0.3cm}
\begin{center}
\caption {Review datasets used in this work.}
\label{Dataset}
\vspace{-0.2cm}
\begin{tabular}{|>{\centering\arraybackslash}m{1.6cm}|>{\centering\arraybackslash}m{1.75cm}|>{\centering\arraybackslash}m{0.9cm}|>{\centering\arraybackslash}m{1.9cm}|}\hline
{Dataset}&{ Reviews (spam\%)}&{Users}&{Business (Resto. \& hotels)} \\ \hline
{Main}&{608,598 (13\%)}&{260,277}&{5,044} \\ \hline
{Review-based}&{62,990 (13\%)}&{48,121}&{3,278} \\ \hline
{Item-based}&{66,841 (34\%)}&{52,453}&{4,588} \\ \hline
{User-based}&{183,963 (19\%)}&{150,278}&{4,568} \\ \hline
{\textcolor[rgb]{0,0,0}{Amazon}}&{\textcolor[rgb]{0,0,0}{8,160 (-)}}&{\textcolor[rgb]{0,0,0}{7685}}&{\textcolor[rgb]{0,0,0}{243}} \\ \hline
\end{tabular}
\end{center}
\vspace{-0.2cm}
\end{table}

\begin{figure*}
\centering
\includegraphics[height=3cm]{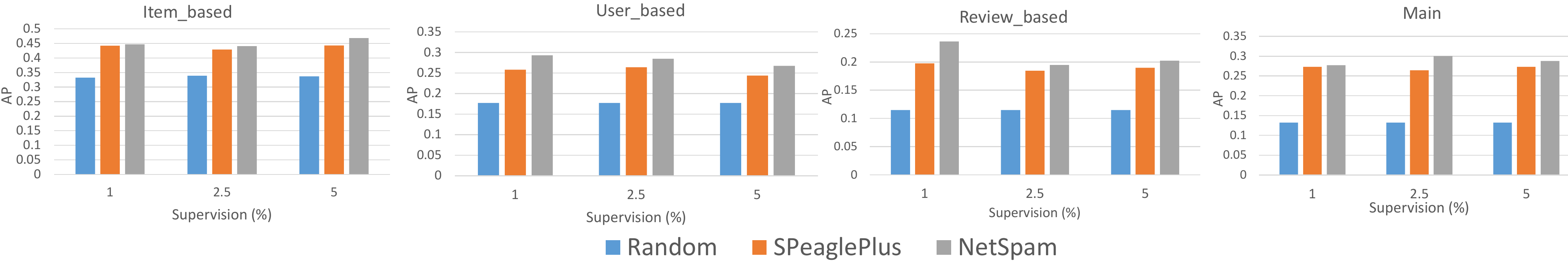}
\vspace{-.24cm}
\caption {\textsl{AP} for Random, SPeaglePlus and \textsl{NetSpam} approaches in different datasets and supervisions (1\%, 2.5\% and 5\%)}
\label{fig:ap-all-features}
\end{figure*}

\begin{figure*}
\centering
\includegraphics[height=2.8cm]{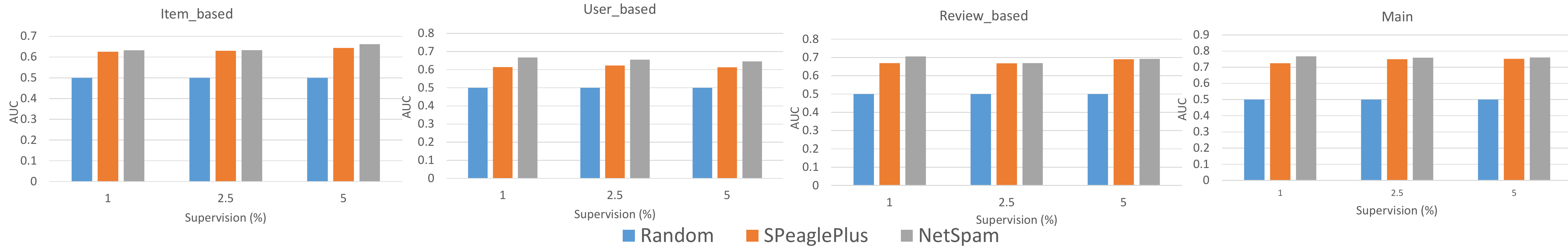}
\vspace{-.2cm}
\caption {\textsl{AUC} for Random, SPeaglePlus and \textsl{NetSpam} approaches in different datasets and supervisions (1\%, 2.5\% and 5\%).}
\label{fig:auc-all-features}
\end{figure*}

\subsection{Evaluation Metrics}
\textcolor[rgb]{0,0,0}{We have used Average Precision (AP) and Area Under the Curve (AUC) as two metrics in our evaluation. %based on features we've just used.
AUC measures accuracy of our ranking based on False Positive Ratio (FPR as y-axis) against True Positive Ratio (TPR as x-axis) and integrate values based on these two measured values.} 
The value of this metric increases as the proposed method performs well in ranking, and vise-versa.
Let \textsl{A} be the list of sorted spam reviews so that $A(i)$ denotes a review sorted on the $i^{th}$ index in $A$. If the number of spam (non-spam) reviews before review in the $j^{th}$ index is equal to $n_j$ and the total number of spam (non-spam) reviews is equal to $f$, then \textsl{TPR} (\textsl{FPR}) for the $j^{th}$
is computed as $\frac{n_j}{f}$.
To calculate the $AUC$, we set $TPR$ values as the \textsl{x-axis} and $FPR$ values on the \textsl{y-axis} and then integrate the area under the curve for the curve that uses their values. We obtain a value for the $AUC$ using:
\begin{equation}
AUC = \sum_{i=2}^{n}(FPR(i) - FPR(i-1))*(TPR(i))
\end{equation}
where $n$ denotes number of reviews. For $AP$ we first need to calculate index of top sorted reviews with spam labels. Let indexes of sorted spam reviews in list $A$ with spam labels in ground truth be like list $I$, then for $AP$ we have:
\begin{equation}
AP = \sum_{i=1}^{n}\frac{i}{I(i)}
\end{equation}

As the first step, two metrics are rank-based which means we can rank the final probabilities. Next we calculate the AP and AUC values based on the reviews' ranking in the final list.

In the most optimum situation, all of the spam reviews are ranked on top of sorted list; In other words, when we sort spam probabilities for reviews, all of the reviews with spam labels are located on top of the list and ranked as the first reviews.
With this assumption we can calculate the AP and AUC values. They are both highly dependent on the number of features. For the learning process, we use different supervisions and we train a set for weight calculation. We also engage these supervisions as fundamental labels for reviews which are chosen as a training set.

\begin{figure*}
\centering
\includegraphics[height=3cm]{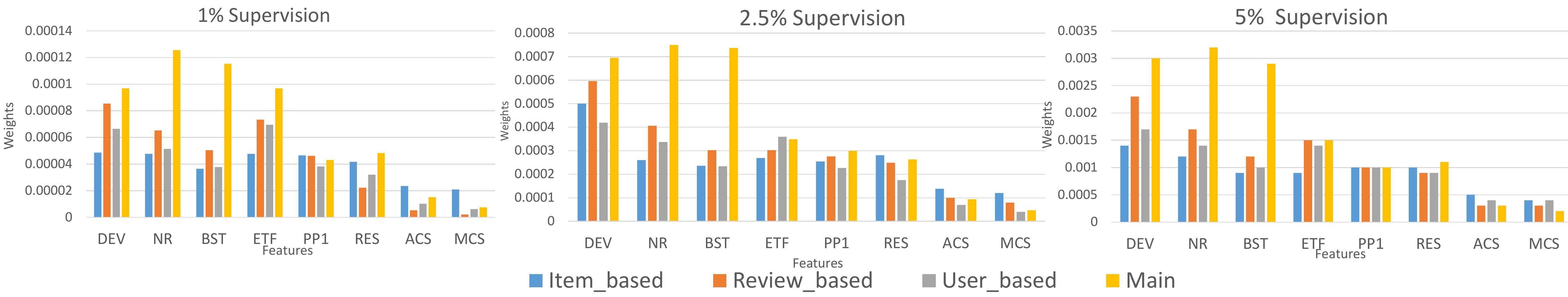}
\vspace{-.2cm}
\caption {Features weights for NetSpam framework on different datasets using different supervisions (1\%, 2.5\% and 5\%).}
\label{fig:weigths-different-features}
\end{figure*}

\subsection{Main Results}
In this section, we evaluate \textsl{NetSpam} from different perspective and compare it with two other approaches, Random approach and SPeaglePlus \cite{Shebuti2015}. 
To compare with the first one, 
we have developed a network in which reviews are connected to each other randomly. Second approach use a well-known graph-based algorithm called as ``LBP" to calculate final labels. 
Our observations show \textsl{NetSpam}, outperforms these existing methods. 
Then analysis on our observation is performed and finally we will examine our framework in unsupervised mode. \textcolor[rgb]{0,0,0}{Lastly, we investigate time complexity of the proposed framework and the impact of camouflage strategy on its performance.}

\subsubsection{Accuracy}
Figures \ref{fig:ap-all-features} and \ref{fig:auc-all-features} present the performance in terms of the \textsl{AP} and \textsl{AUC}. As it's shown in all of the four datasets \textsl{NetSpam} outperforms SPeaglePlus specially when number of features increase. 
In addition different supervisions have no considerable effect on the metric values neither on \textsl{NetSpam} nor SPeaglePlus. 
Results also show the datasets with higher percentage of spam reviews have better performance because when fraction of spam reviews in a certain dataset increases, probability for a review to be a spam review increases and as a result more spam reviews will be labeled as spam reviews and in the result of \textsl{AP} measure which is highly dependent on spam percentage in a dataset.
On the other hand, \textsl{AUC} measure does not fluctuate too much, because this metric is not dependent on spam reviews percentage in dataset, but on the final sorted list which is calculated based on the final spam probability.

\begin{table*}
\begin{center}
\caption {Weights of all features (with \textbf{5\% data as train set}); features are ranked based on their overall average weights.}
\vspace{-.2cm}
\label{features-all-weights}
\begin{tabular}{| c | c  | c | c | c | c| c|c| c |}\hline
{Dataset - Weights} & {DEV} & {NR} & {ETF} & {BST} & {RES} & {PP1} & {ACS} & {MCS}
\\ \hline
{Main}&{0.0029}&{0.0032}&{0.0015}&{0.0029}&{0.0010}&{0.0011}&{0.0003} & {0.0002}\\
{Review-based}&{0.0023}&{0.0017}&{0.0017}&{0.0015}&{0.0010}&{0.0009}&{0.0004} &{0.0003}\\
{Item-based}&{0.0010}&{0.0012}&{0.0009}&{0.0009}&{0.0010}&{0.0010}&{0.0004} &{0.0003}\\
{User-based}&{0.0017}&{0.0014}&{0.0014}&{0.0010}&{0.0010}&{0.0009}&{0.0005} &{0.0004}
\\ \hline
\end{tabular}
\end{center}
\end{table*}

\subsubsection{Feature Weights Analysis}
\label{subsec:Unsupervised}
Next we discuss about features weights and their involvement to determine spamicity. First we inspect how much \textsl{AP} and \textsl{AUC} are dependent on variable number of features. Then we show these metrics are different for the four feature types explained before (RB, UB, RL and UL). To show how much our work on weights calculation is effective, first we have simulated framework on several run with whole features and used most weighted features to find out best combination which gives us the best results. Finally, we found which category is most effective category among those listed in Table \ref{Feature-list}. 

\textbf{\textit{Dataset Impression on Spam Detection:}}
As we explained previously, different datasets yield different results based on their contents. For all datasets and most weighted features, there is a certain sequence for features weights. As is shown in Fig. \ref{fig:weigths-different-features} for four datasets, in almost all of them, features for the Main dataset have more weights and features for Review-based dataset stand in the second position. Third position belongs to User-based dataset and finally Item-based dataset has the minimum weights (for at least the four features with most weights).

\textbf{\textit{Features Weights Importance:}}
\label{feature-importance}
As shown in Table \ref{features-all-weights}, there are couple of features which are more weighted than others. Combination of these features can be a good hint for obtaining better performance.
The results of the Main dataset show all the four behavioral features are ranked as first features in the final overall weights. In addition, as shown in the Review-based as well as other two datasets, $DEV$ is the most weighted feature. This is also same for our second most weighted feature, $NR$. From the third feature to the last feature there are different order for the mentioned features. The third feature for both datasets User-based and Review-based is same, $ETF$, while for the other dataset, Item-based, $PP1$ is at rank 3. Going further, we see in the Review-based dataset all four most weighted features are behavioral-based features which shows how much this type of features are important in detecting spams as acknowledged by other works as well \cite{Shebuti2015, Mukherjee2013-2}.

\begin{figure}
\vspace{-.2cm}
\centering
\includegraphics[height=5cm]{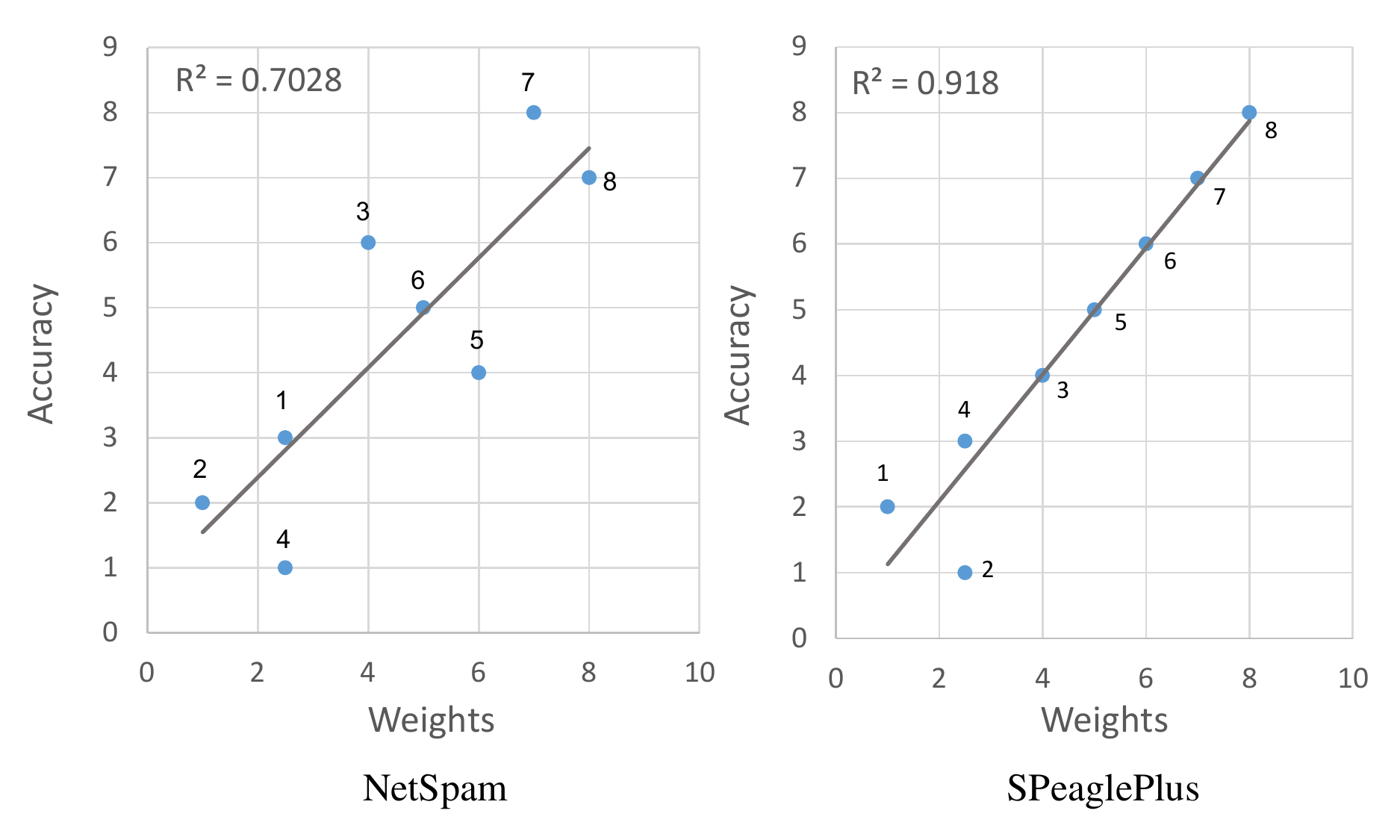}
\vspace{-.3cm}
\caption {Regression graph of features vs. accuracy (with \textbf{5\% data as train set}) for \textsl{Main} dataset. (see Table \ref{MetaPaths} for numbers)
}
\label{fig:regression-semi}
\vspace{-.3cm}
\end{figure}

As we can see in Fig. \ref{fig:regression-semi}, there is a strong correlation between features weights and the accuracy.
For the Main dataset we can see this correlation is much more obvious and also applicable. Calculating weights using \textsl{NetSpam} help us to understand how much a feature is effective in detecting spam reviews; since as much as their weights increase two metrics including AP and AUC also increase respectively and therefore our framework can be helpful in detecting spam reviews based on features importance.

\begin{figure*}
\centering
\includegraphics[height=4cm]{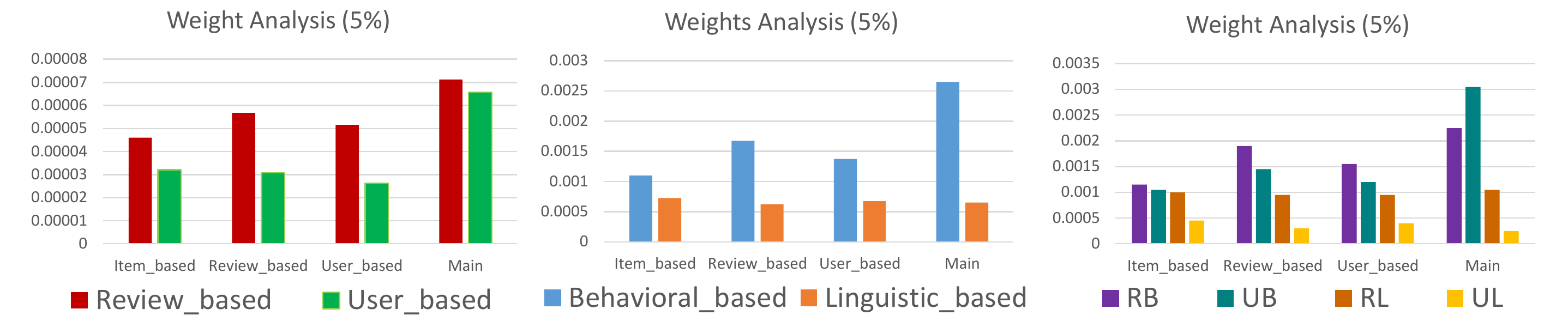}
\vspace{-.6cm}
\caption {Features weights for different features categories (RB, UB, RL and UL) with 5\% supervision, on different datasets.}
\label{fig:features-category-analysis}
\vspace{-.3cm}
\end{figure*}

The observations indicate larger datasets yield better correlation between features weights and also its accuracy in term of AP. Since we need to know each feature rank and importance we use Spearman's rank correlation for our work. In this experience our main dataset has correlation value equal to 0.838 (p-value=0.009), while this value for our next dataset, User-based one, is equal to 0.715 (p-value = 0.046). As much as the size of dataset gets smaller in the experiment, this value drops. This problem is more obvious in Item and Review-based datasets. For Item-based dataset, correlation value is 0.458 which is low, because sampling Item-based dataset needs Item-based features. The features are identical to each item and are similar to user-based features.
Finally the obtained results for our smallest dataset is satisfying, because final results considering AP show a correlation near to 0.683 between weights and accuracy (similar results for SPeaglePlus as well). Weights and accuracy (in terms of AP) are completely correlated. We observed values 0.958 (p-value=0.0001), 0.764 (p=0.0274), 0.711 (p=0.0481) and 0.874 (p=0.0045) for the Main, User-based, Item-based and Review-based datasets, respectively.
This result shows using weight calculation method and considering metapath concept can be effective in determining the importance of features. Similar result for SPeaglePlus also shows our weights calculation method
can be generalized to other frameworks and can be used as a main component for finding each feature weight.

Our results also indicate feature weights are completely dependent on datasets, considering this fact two most important features in all datasets are same features.
This means except the first two features, other features weights are highly variable regrading to dataset used for extracting weights of features.

\textbf{\textit{Features Category Analysis:}}
\label{category-analysis}
As shown in Fig. \ref{fig:features-category-analysis} there are four categories with different weights average which is very important, specially in determining which feature is more appropriate for spotting spam reviews (refer to Sec. \ref{feature-importance}). Since results for different supervision are similar we have just presented the results for 5\% supervision. We have analyzed features based on their categories and obtained results in all datasets show that Behavioral-based features have better weights than linguistic ones which is confirmed by \cite{Mukherjee2013-1} and \cite{Shebuti2015}. Analysis on separate views shows that review-based features have higher weights which leads to better performance. It is worth to mention that none of previous works have investigated this before.
Same analysis on the \textsl{Main} dataset shows equal importance of both category in finding spams. \textcolor[rgb]{0,0,0}{On the Other hand, in the first three dataset from Table \ref{Feature-list}, RB has better weights (a bit difference in comparison with RU), which means this category yields better performance than other categories for spotting spam reviews. Differently, for \textsl{Main} dataset UB categories has better weights and has better performance than RU category and also other categories, in all datasets behavioral-based features yield better performance with any supervision.}

\subsubsection{Unsupervised Method}
\label{sec:unsupervised}
One of the achievement in this study is that even without using a train set, we can still find the best set of features which yield to the best performance. As it is explained in Sec. \ref{sec:review-network-section}, in unsupervised approach special formulation is used to calculate fundamental labels and next these labels are used to calculate the features' weight and finally review labels. As shown in Fig. \ref{fig:regression-un}, our observations show there is a good correlation in the Main dataset in which for \textsl{NetSpam} it is equal to 0.78 (p-value=0.0208) and for SPeaglePlus this value reach 0.90 (p=0.0021). As another example for user-based dataset there is a correlation equal to 0.93 (p=0.0006) for \textsl{NetSpam}, while for SPeagle this value is equal to 0.89 (p=0.0024). This observation indicates \textsl{NetSpam} can prioritize features for both frameworks. 
\textcolor[rgb]{0,0,0}{Table \ref{features-all-weights-unsupervised} demonstrates that there is certain sequence in feature weights and it means in spam detection problems, spammers and spam reviews have common behaviors, no matter what social network they are writing the review for: Amazon or Yelp. For all of them, $DEV$ is most weighted features, followed by $NR$, $ETF$ and $BST$.}

\begin{table*}
\begin{center}
\caption {Weights of all features (using \textbf{unsupervised} approach); features are ranked based on their overall average weights.}
\vspace{-.2cm}
\label{features-all-weights-unsupervised}
\begin{tabular}{| c | c  | c | c | c | c| c|c| c |}\hline
{Dataset - Weights} & {DEV} & {NR} & {ETF} & {BST} & {RES} & {PP1} & {ACS} & {MCS}
\\ \hline
{\textsl{Main}}&{0.0029}&{0.0550}&{0.0484}&{0.0445}&{0.0379}&{0.0329}&{0.0321} & {0.0314}\\
{\textsl{Review-based}}&{0.0626}&{0.0510}&{0.0477}&{0.0376}&{0.0355}&{0.0346}&{0.0349} &{0.0340}\\
{\textsl{Item-based}}&{0.0638}&{0.0510}&{0.0501}&{0.0395}&{0.0388}&{0.0383}&{0.0374} &{0.0366}\\
{\textsl{User-based}}&{0.0630}&{0.0514}&{0.0494}&{0.0380}&{0.0373}&{0.0377}&{0.0367} &{0.0367}\\
{\textsl{\textcolor[rgb]{0,0,0}{Amazon}}}&{\textcolor[rgb]{0,0,0}{0.1102}}&{\textcolor[rgb]{0,0,0}{0.0897}}&{\textcolor[rgb]{0,0,0}{0.0746}}&{\textcolor[rgb]{0,0,0}{0.0689}}&{\textcolor[rgb]{0,0,0}{0.0675}}&{\textcolor[rgb]{0,0,0}{0.0624}}&{\textcolor[rgb]{0,0,0}{0.0342}} &{\textcolor[rgb]{0,0,0}{0.0297}}
\\ \hline
\end{tabular}
\end{center}
\vspace{-.24cm}
\end{table*}

\begin{center}
\begin{figure}
\vspace{-.2cm}
\centering
\includegraphics[height=5cm]{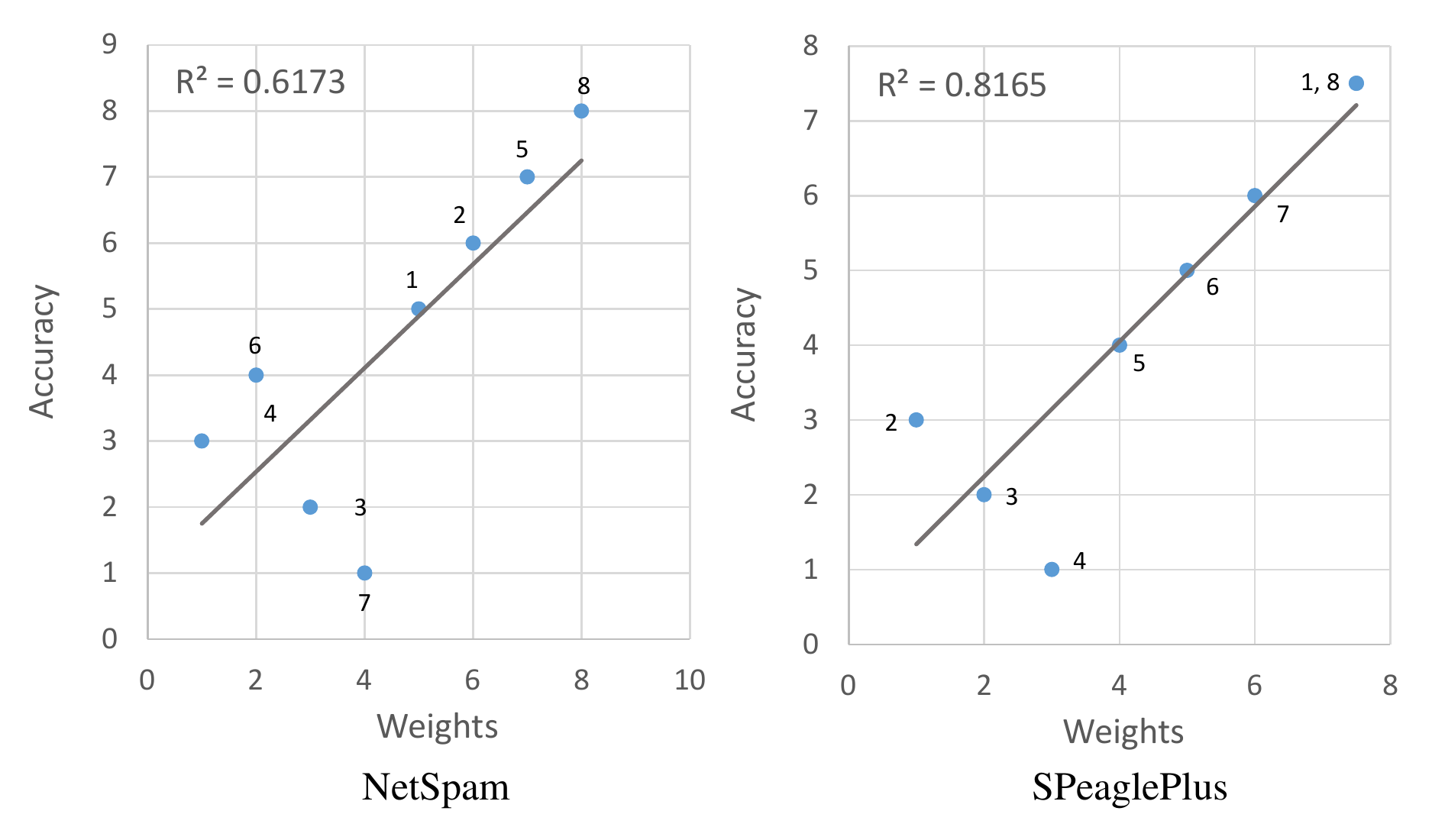}
\vspace{-.3cm}
\caption {Regression graph of features vs. accuracy (\textbf{unsupervised}) for \textsl{Main} dataset. (see Table \ref{MetaPaths} for numbers)
}
\label{fig:regression-un}
\end{figure}
\vspace{-.25cm}
\end{center}

\textcolor[rgb]{0,0,0}{
\vspace{-.8cm}
\subsubsection{Time Complexity}
If we consider the \textsl{Main} dataset as input to our framework, time complexity with these circumstances is equal to \textit{$O(e^2m)$} where \textit{e} is number of edges in created network or reviews number. It means we need to check if there is a metapath between a certain node (review) with other nodes which is \textit{$O(e^2)$} and this checking must be repeated for very feature. So, our time complexity for offline mode in which we give the \textsl{Main} dataset to framework and calculate spamicity of whole reviews, is \textit{$O(e^2m)$} where \textit{$m$} is number of features.}
\textcolor[rgb]{0,0,0}{
In online mode, %where we have a network of reviews in our framework,
a review is given to NetSpam to see whether it is spam or not, we need to check if there is a metapath between given review with other reviews, which is in \textit{$O(e)$}, and like offline mode it has to be repeated for every feature and every value. Therefore the complexity is \textit{$O(em)$}.
}

\textcolor[rgb]{0,0,0}{
\subsubsection{The Impact of Camouflage Strategy}
One of the challenges that spam detection approaches face is that spammers often write non-spam reviews to hide their true identity known as camouflage. For example they write positive reviews for good restaurant or negative reviews for low-quality ones; hence every spam detector system fails to identify this kind of spammers or at least has some trouble to spot them. In the previous studies, there are different approaches for handling this problem. For example, in \cite{Shebuti2015}, the authors assumes there is always a little probability that a good review written by a spammer and put this assumption in its compatibility matrix.
In this study, we tried to handle this problem by using weighted metapaths. In particular, we assume that even if a review has a very little value for a certain feature, it is considered in feature weights calculation. Therefore, instead of considering metapaths as binary concepts, we take 20 values which denoted as $s$. Indeed, if there is a camouflage its affection will be reduced. As we explained in Section \ref{metapath-creation} in such problems it is better to propose a fuzzy framework, rather than using a bipolar values $(0,1)$.
}

%%%%%%%%%%%%%%%%%%%%%%%%%%%%%%%%%%%%%%%%%%%
%\input{2_Related_work}
\section{Related Works}
\label{sec:Related_Work}
In the last decade, a great number of research studies focus on the problem of spotting spammers and spam reviews. However, since the problem is non-trivial and challenging, it remains far from fully solved.
We can summarize our discussion about previous studies in three following categories.
 
\subsection{Linguistic-based Methods} This approach extract linguistic-based features to find spam reviews.
Feng \textit{et al.} \cite{Feng2012} use $unigram$, $bigram$ and their composition.
Other studies \cite{Xu2014}, \cite{Li2011}, \cite{Lim2010} use other features like pairwise features (features between two reviews; e.g. content similarity), percentage of CAPITAL words in a reviews for finding spam reviews.
\textcolor[rgb]{0,0,0}{Lai \textit{et al.} in \cite{Lai2011} use a probabilistic language modeling to spot spam. This study demonstrates that 2\% of reviews written on business websites are actually spam.}

\subsection{Behavior-based Methods} Approaches in this group almost use reviews metadata to extract features; those which are normal pattern of a reviewer behaviors.
Feng \textit{et al.} in \cite{Feng2012b} focus on distribution of spammers rating on different products and traces them.
In \cite{Jindal2008}, Jindal \textit{et. al} extract 36 behavioral features and use a supervised method to find spammers on Amazon and \cite{Jindal2010} indicates behavioral features show spammers' identity better than linguistic ones.
\textcolor[rgb]{0,0,0}{ Xue \textit{et al.} in \cite{Xue2015} use rate deviation of a specific user and use a trust-aware model to find the relationship between users for calculating final spamicity score.}
Minnich \textit{et al.} in \cite{Minnich2015} use temporal and location features of users to find unusual behavior of spammers.
Li \textit{et al.} in \cite{Li2014} use some basic features (e.g polarity of reviews) and then run a HNC (Heterogeneous Network Classifier) to find final labels on Dianpings dataset.
Mukherjee \textit{et al.} in \cite{Mukherjee2013-1} almost engage behavioral features like  rate deviation, extremity and etc.
Xie \textit{et al.} in \cite{Xie2012} also use a temporal pattern (time window) to find singleton reviews (reviews written just once) on Amazon.
Luca \textit{et al.} in \cite{Luca2016} use behavioral features to show increasing competition between companies leads to very large expansion of spam reviews on products.

Crawford \textit{et al.} in \cite{Michael2016} indicates using different classification approach need different number of features to attain desired performance and propose approaches which use fewer features to attain that performance and hence recommend to improve their performance while they use fewer features which leads them to have better complexity. With this perspective our framework is arguable. This study shows using different approaches in classification yield different performance in terms of different metrics.

\subsection{Graph-based Methods}
Studies in this group aim to make a graph between users, reviews and items and use connections in the graph and also some network-based algorithms to rank or label reviews (as spam or genuine) and users (as spammer or honest). Akoglu \textit{et al.} in \cite{Akoglu2013} use a network-based algorithm known as LBP (Loopy Belief Propagation)
in linearly scalable iterations related to number of edges to find final probabilities for different components in network.
Fei \textit{et al.} in \cite{Fei2013} also use same algorithm (LBP), and utilize burstiness of each review to find spammers and spam reviews on Amazon. Li \textit{et al.} in \cite{Li2014} build a graph of users, reviews, users IP and indicates users with same IP have same labels, for example if a user with multiple different account and same IP writes some reviews, they are supposed to have same label. Wang \textit{et al.} in \cite{Wang2011} also create a network of users, reviews and items and use basic assumptions (for example a reviewer is more trustworthy if he/she writes more honest reviews) and label reviews. Wahyuni in \cite{Eka2016} proposes a hybrid method for spam detection using an algorithm called ICF++ which is an extension to ICF of \cite{Wang2011} in which just review rating are used to find spam detection. This work use also sentiment analysis to achieve better accuracy in particular.

Deeper analysis on literature show that behavioral features work better than linguistic ones in term of accuracy they yield. There is a good explanation for that; in general, spammers tend to hide their identity for security reasons. Therefore they are hardly recognized by reviews they write about products, but their behavior is still unusual, no matter what language they are writing. In result, researchers combined both feature types to increase accuracy of spam detection. The fact that adding each feature is a time consuming process, this is where feature importance is useful. Based on our knowledge, there is no previous method which engage importance of features (known as weights in our proposed framework; \textsl{NetSpam}) in the classification step. By using these weights, on one hand we involve features importance in calculating final labels and hence accuracy of \textsl{NetSpam} increase, gradually. On the other hand we can determine which feature can provide better performance in term of their involvement in connecting spam reviews (in proposed network).

%%%%%%%%%%%%%%%%%%%%%%%%%%%%%%%%%%%%%%%%%%%
%\input{6_Conclusion}
\section{Conclusion}
\label{sec:Conclusion}
This study introduces a novel spam detection framework namely \textsl{NetSpam} based on a metapath concept as well as a new graph-based method to label reviews relying on a rank-based labeling approach. 
The performance of the proposed framework is evaluated by using two real-world labeled datasets of Yelp and Amazon websites. 
Our observations show that calculated weights by using this metapath concept can be very effective in identifying spam reviews and leads to a better performance. 
In addition, we found that even without a train set, \textsl{NetSpam} can calculate the importance of each feature and it yields better performance in the features' addition process, and performs better than previous works, with only a small number of features.
Moreover, after defining four main categories for features our observations show that the reviews-behavioral category performs better than other categories, in terms of AP, AUC as well as in the calculated weights. The results also confirm that using different supervisions, similar to the semi-supervised method, have no noticeable effect on determining most of the weighted features, just as in different datasets.

For future work, metapath concept can be applied to other problems in this field. For example, similar framework can be used to find spammer communities. For finding community, reviews can be connected through group spammer features (such as the proposed feature in \cite{Mukherjee2012}) and reviews with highest similarity based on metapth concept are known as communities. In addition, utilizing the product features is an interesting future work on this study as we used features more related to spotting spammers and spam reviews.
Moreover, while single networks has received considerable attention from various disciplines for over a decade, information diffusion and content sharing in multilayer networks is still a young research \cite{salehi2015spreading}. Addressing the problem of spam detection in such networks can be considered as a new research line in this field.

\section{Acknowledgment}
\vspace{-0.15cm}
This work is partially supported by Iran National Science Foundation (INSF) (Grant No. 94017889).
\vspace{-0.2cm}

%%%%%%%%%%%%%%%%%%%%%%%%%%%%%%%%%%%%%%%%%%

\end{document}